\documentclass[sigconf]{acmart}
\usepackage{soul,color}
\soulregister\cite7 % 针对\cite命令
\usepackage{threeparttable} 
\AtBeginDocument{%
  \providecommand\BibTeX{{%
    \normalfont B\kern-0.5em{\scshape i\kern-0.25em b}\kern-0.8em\TeX}}}

\copyrightyear{2021}
\acmYear{2021}
\setcopyright{acmlicensed}\acmConference[CHI '21]{CHI Conference on Human Factors in Computing Systems}{May 8--13, 2021}{Yokohama, Japan} \acmBooktitle{CHI Conference on Human Factors in Computing Systems (CHI '21), May 8--13, 2021, Yokohama, Japan}
\acmPrice{15.00}
\acmDOI{10.1145/3411764.3445763}
\acmISBN{978-1-4503-8096-6/21/05}

\begin{document}

%%
%% The "title" command has an optional parameter,
%% allowing the author to define a "short title" to be used in page headers.
\title{Data Engagement Reconsidered: A Study of Automatic Stress Tracking Technology in Use}

%%
%% The "author" command and its associated commands are used to define
%% the authors and their affiliations.
%% Of note is the shared affiliation of the first two authors, and the
%% "authornote" and "authornotemark" commands
%% used to denote shared contribution to the research.

\author{Xianghua Ding}
\email{dingx@fudan.edu.cn}
\affiliation{%
  \institution{Shanghai Key Laboratory of Data Science, School of Computer Science}
  \institution{Fudan University}
  \streetaddress{Songhu Road 2005}
  \city{Shanghai}
  \country{China}
}

\author{Shuhan Wei}
\email{18210240205@fudan.edu.cn}
\affiliation{%
  \institution{School of Computer Science}
  \institution{Fudan University}
  \streetaddress{Songhu Road 2005}
  \city{Shanghai}
  \country{China}
}

\author{Xinning Gui}
\email{xinninggui@psu.edu}
\affiliation{%
  \institution{College of Information Sciences and Technology}
  \institution{Pennsylvania State University}
  \streetaddress{University Park, PA 16802}
%   \city{Philadelphia}
%   \state{Pennsylvania}
  \country{USA}
}

\author{Ning Gu}
\email{ninggu@fudan.edu.cn}
\affiliation{%
  \institution{School of Computer Science}
  \institution{Fudan University}
  \streetaddress{Songhu Road 2005}
  \city{Shanghai}
  \country{China}
}

\author{Peng Zhang}
\email{zhpll@126.com}
\affiliation{%
  \institution{School of Computer Science}
  \institution{Fudan University}
  \streetaddress{Songhu Road 2005}
  \city{Shanghai}
  \country{China}
}

%%
%% By default, the full list of authors will be used in the page
%% headers. Often, this list is too long, and will overlap
%% other information printed in the page headers. This command allows
%% the author to define a more concise list
%% of authors' names for this purpose.
\renewcommand{\shortauthors}{Xianghua and Shuhan, et al.}

%%
%% The abstract is a short summary of the work to be presented in the
%% article.
\begin{abstract}
In today's fast-paced world, stress has become a growing health concern. While more automatic stress tracking technologies have recently become available on wearable or mobile devices, there is still a limited understanding of how they are actually used in everyday life. This paper presents an empirical study of automatic stress-tracking technologies in use in China, based on semi-structured interviews with 17 users. The study highlights three challenges of stress-tracking data engagement that prevent effective technology usage: the lack of immediate awareness, the lack of pre-required knowledge, and the lack of corresponding communal support. Drawing on the stress-tracking practices uncovered in the study, we bring these issues to the fore, and unpack  assumptions embedded in related works on self-tracking and how data engagement is approached. We end by calling for a reconsideration of data engagement as part of self-tracking practices with technologies rather than simply looking at the user interface.

\end{abstract}

%%
%% The code below is generated by the tool at http://dl.acm.org/ccs.cfm.
%% Please copy and paste the code instead of the example below.
%%


%%
%% Keywords. The author(s) should pick words that accurately describe
%% the work being presented. Separate the keywords with commas.

%% A "teaser" image appears between the author and affiliation
%% information and the body of the document, and typically spans the
%% page.
% \begin{teaserfigure}
%   \includegraphics[width=\textwidth]{sampleteaser}
%   \caption{Seattle Mariners at Spring Training, 2010.}
%   \Description{Enjoying the baseball game from the third-base
%   seats. Ichiro Suzuki preparing to bat.}
%   \label{fig:teaser}
% \end{teaserfigure}

%%
%% This command processes the author and affiliation and title
%% information and builds the first part of the formatted document.
\maketitle

\section{Introduction}
In today's fast-paced and hectic world, stress is a growing health concern. It is not just that too much stress can reduce study and work efficiency, but that stress has been closely linked to psychological illness. Previous research has found a strong association between stress and depression \cite{hammen2005stress}, and long term exposure to high levels of stress can negatively impact our well being \cite{rosengren2004association,matthews2002chronic,chandola2006chronic}, leading to various physical diseases, such as hypertension \cite{Pickering2001Mental}, cardiovascular disease \cite{Pickering2001Mental}, infectious illnesses \cite{steptoe1991invited} and even cancer \cite{cohen2007psychological}. Therefore, the awareness and effective management of stress is of significant importance to the management of health.

In recent years, stress-tracking technologies that can automatically detect and collect stress data during the day have become commercially available. More and more wearable products have automatic stress tracking features embedded, such as smart bracelets and watches made by Huawei \cite{huawei}, Garmin \cite{garmin}, and Samsung \cite{samsung}. There are also products that are dedicated to stress-tracking, such as Healbe Gobe2 \cite{gobe2}, Wellbe \cite{wellbe}, Bellabeat Leaf Urban \cite{bellabeat}, and Spire Stone \cite{spire}. These products, by automatically detecting an individual's stress level, often combined with features to help with relaxation, offer the potential to help with an awareness and an effective management of stress on a daily basis. 

With more mature stress-tracking technologies on the market, however, it is still unclear how these technologies actually work in practice. In HCI and related fields, there have been many  studies on stress, but they've primarily focused  on innovative approaches to automatic stress tracking \cite{hovsepian2015cstress, bogomolov2014daily,li2018photoplethysmography,lu2012stresssense} or designs that can help stress relief \cite{cochrane2019reconnecting, paredes2011calmmenow, yu2017stresstree}; little attention has been paid to how people use the automatic stress-tracking technologies. In addition, although there has been extensive work on the use of self-tracking technologies, also known as Personal Informatics (PI) \cite{li2010stage}, or quantified self \cite{choe2014understanding}, that are designed to track various aspects of our lives, such as steps, mood, sleep, and heart rate \cite{lin2006fish, caldeira2017mobile, ravichandran2017making, harrison2015activity, lee2018mindnavigator}, research on the use of stress-tracking technologies in everyday lives in particular has been rare. Yet, stress-tracking has distinct characteristics that deserve their own attention. Stress as a measurement is not as straightforward as steps \cite{bassett2017step}, or heart rate that can be directly quantified with counting, and is less objective and more complex. Furthermore, stress involves not only psychological and emotional responses (such as anxiety, anger, sadness \cite{lazarus1991emotion}, fear, and frustration \cite{butler1993definitions}), but also physiological and bodily reactions. As \cite{kelley2017self} points  out, self-tracking for daily stress has unique challenges because stress is highly subjective and involves social and environmental factors. Thus, the research question we would like to answer is this: how do people encounter and use the automatic stress-tracking technologies that have become available in more and more wearable devices in everyday life?

To answer it, we conducted a qualitative study to understand automatic stress-tracking technology in use. We recruited 17 participants in China who used automatic stress-tracking technologies and conducted semi-structured interviews with them. The study highlights a number of challenges associated with users' stress-tracking data engagement, including the lack of immediate awareness of relevant data, the lack of pre-required knowledge, domain and technical, as well as the lack of corresponding communities of practice.  Many of these challenges are associated  with how the automatic stress-tracking technology is adopted and designed, how the stress data is encountered, and how our users are socially situated. This study unpacks some of the data engagement assumptions embedded in the related work on self-tracking technologies.

The contribution of this paper is an empirical study on the use of automatic stress tracking in practice, and a more nuanced understanding of data engagement with self-tracking technologies. In the paper below, we will first give background information on stress and stress-tracking technologies and  review related works on stress and self-tracking data engagement. We will then present our study and the findings, and discuss how the focus of stress-tracking technologies brings to the fore some of the issues of data engagement with self-tracking technologies in general.

\section{Background}

\subsection{What is stress?}
While the term ``stress'' is pervasively used, there has never been a unified definition of it. Broadly speaking, stress has been mainly examined in two ways, psychologically and physiologically, with the former focusing on psychological feelings and the perception of stress and the latter referring to the bodily response to external events. 

%is defined as the stress and coping theory \cite{lazarus1984stress} is commonly used for stress definition, it holds that...physical impairments (e.g. depression, cancer), this kind of stress is called ``distress'' (negative stress). But there is another type of stress called ``eustress'' (positive stress), which is often overlooked. Most of the papers on stress are about general stress or distress, but less about eustress \cite{kupriyanov2014eustress}. In 1976, Selye drew first health-centered distinction between distress and eustress \cite{selye1976stress}.  As opposed to distress...is associated with positive feelings (e.g. gratitude, hope，good will, enjoy the work \cite{heikkila2015potential}) and healthy physical states (e.g. enhance immune system \cite{lazarus1993psychological}). Under eustress

In psychology, stress refers to the feelings and perception of pressure. It holds that ``stress occurs when a person perceives the demands of an environment stimuli to be greater than their ability to meet, mitigate, or alter those demands'' \cite{lazarus1985stress}. Stress is as such perceived as a subjective concept, and in psychology, self-reporting is usually used to detect it . While most associate stress  with negative feelings, such as fear and anxiety, stress can also be positive and beneficial. Unlike  negative stress or ``distress'', with  positive stress or ``eustress'' , people appraise a situation to be  challenging and non- threatening \cite{folkman2000stress} and have the confidence to solve it. One study found an inverted u-shaped relationship between stress and performance; in other words, stress is beneficial to performance until an optimal level, and then performance starts to decrease \cite{le2003eustress}. 

In the medical field, the term ``stress'' is defined physiologically as ``the non-specific responses of the body to any demand for change'' \cite{selye1965stress}. When people encounter threats or challenges, the body will have corresponding reactions, which are generated by the autonomic nervous system. The autonomic nervous system (ANS) is comprised of the sympathetic nervous system (SNS) and the parasympathetic nervous system (PNS). The SNS is responsible for mobilizing the body's resources to deal with stressful events, in what is called the ``fight-or-flight'' response, and brings with it a series of physiological reactions, such as an increase in heart rate,  and respiration and sweat gland activity  \cite{sun2010activity}. The PNS is mainly active during  periods of relaxation and recovery. 

% ``Stressors'' and ``coping'' are two important concepts in stress management. Carlesworth \& Nathan have defined stressors as the external demands of life or the internal attitude and thoughts that require us to adopt????\cite{charlesworth2004stress}. In this view, there are two kinds of stressors: external stressors (eg. work deadlines, economic problems, traffic jams) and internal stressors (eg. negative self-evaluation, unrealistic expectations). A widely used definition of ``coping'' has been proposed by Folkman and Lazarus as ``the cognitive and behavioral efforts made to master, tolerate, or reduce external and internal demands and conflicts'' \cite{folkman1980analysis}. Coping is further divided into two types, problem-focused and emotion-focused \cite{folkman1980analysis, lazarus1984stress}. Problem-focused coping refers to solving problems or eliminating stressors, while emotion-focused coping is  the adjustment of  emotions or thoughts, rather than the situation itself.

When people talk about stress in everyday life, they are usually referring to psychological or subjective feelings, such as tension, anxiety, and fear, which they frequently associate with an external event, like  an upcoming  exam or deadline at work. While more people are starting to realize the impact of long-term stress on their health, the way  the nervous system physiologically reacts to stress and  the distinction between psychological and physiological stress, are not yet a part of most people's everyday understanding. 

\subsection{Stress-Tracking Technology}
Today, there are wearable commercial stress-tracking  devices available on the market that can automatically detect stress including  general products, such as smart bracelets and  watches from Huawei \cite{huawei}, Garmin \cite{garmin} and Samsung \cite{samsung} with embedded stress -tracking features, as well as other products like Healbe Gobe2 \cite{gobe2}, Wellbe \cite{wellbe}, Bellabeat Leaf Urban \cite{bellabeat}, and Spire Stone \cite{spire}  specializing in stress-tracking. The products by Huawei, Garmin, and Samsung detect stress based on an analysis of Heart Rate Variability (HRV)  collected by an embedded optical heart rate sensor. Some devices, such as the Huawei (including Honor) watch, ask users to fill out a stress questionnaire when they first start using the stress-tracking function. 

Heart Rate Variability (HRV), defined as the variation over time of the period between consecutive heartbeats (R-R intervals) \cite{acharya2006heart}, has been proven to be a reliable indicator of   ANS activity \cite{malik1996heart} and can be used as an objective assessment of stress \cite{kim2018stress}. HRV is widely used for stress detection with both laboratory stressors ( arithmetic problems \cite{gandhi2015mental},the  Stroop Color Word Test, highly paced video games \cite{castaldo2017extent}) and real life stressors (university examinations \cite{melillo2011nonlinear}, speeches \cite{aguiar2013speech}, driving \cite{munla2015driver}). An accurate HRV is usually obtained from an Electrocardiogram (ECG) sensor, which needs to attached by electrodes or chest straps directly to the body. A less invasive and more comfortable alternative  is Photoplethysmography (PPG), which can be embedded into a phone camera, ring, or smart wristband device. The Pulse rate variability (PRV) extracted from the PPG has also been proven to be an effective surrogate for HRV for stress detection \cite{lyu2015measuring, lin2014comparison}. In fact, in many  wearable commercial products, PRV is directly referred to as HRV. In this paper, we will not distinguish between HRV and PRV , as we will mainly be studying wearable commercial stress-tracking products. 

%Besides physiological signals, some products also use contextual information for stress detection or prediction. For example, Healbe Gobe2 detects stress based on HRV, sleep quality data, and anthropometrical parameters like weight, height, sex, and age. WellBe determines one's stress and calmness levels based on HRV and  contextual information including time, location and the people met throughout the day. Bellabeat Leaf Urban is smart jewelry designed specifically for women's health, and can predict stress based on sleep, exercise, meditation, and menstrual cycle.  Spire Stone can be clipped to a  bra or waistband to measure one's breathing and classify it as calm, tense, or focused.

\begin{figure*}
    \centering
    \begin{minipage}[t]{8cm}
        \centering
        \includegraphics[width=5cm]{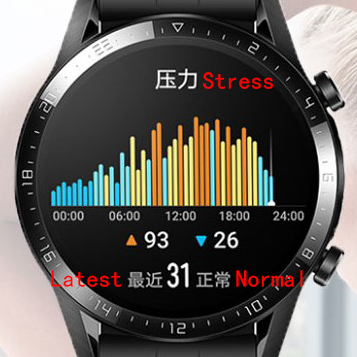}
        \caption{The stress interface on Watch GT}
        \Description{The stress interface on Watch GT, with four different colours represent four different stress levels.}
        \label{fig:WatchGT}
    \end{minipage}
    \begin{minipage}[t]{8cm}
        \centering
        \includegraphics[width=5cm]{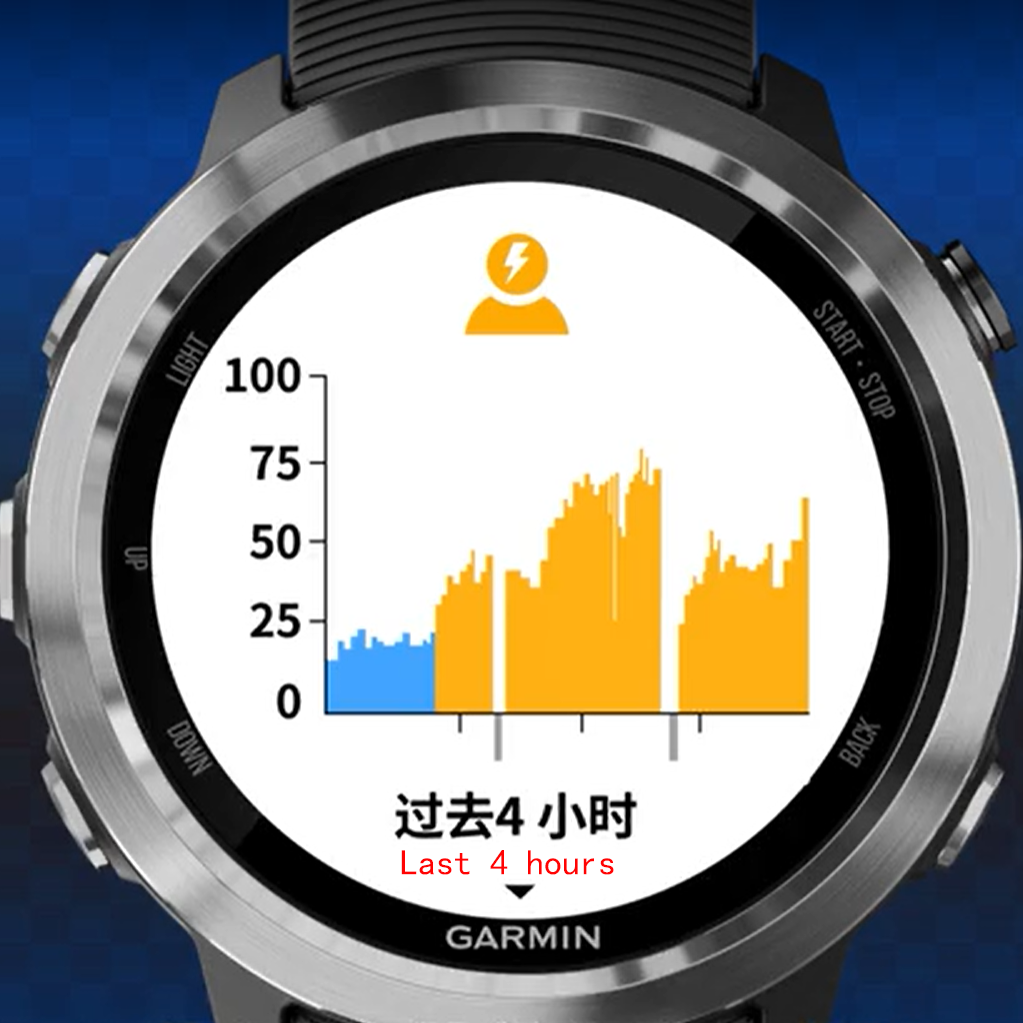}
        \caption{The stress interface on Garmin}
        \Description{The stress interface on Garmin, with blue and yellow represent resting and stress respectively.}
        \label{fig:Garmin}
    \end{minipage}
\end{figure*}

 These products also commonly provide visualizations to help users engage with the quantified stress data.  For example, on Huawei's stress-tracking products, stress value, in range of 1-99, is displayed every 30 minutes, and is divided into four levels (1-29 as relaxed, 30-59 as normal, 60-79 as medium, 80-99 as high), which are shown in bars with different corresponding colours (sky blue, light blue, yellow and orange), as shown in Fig.~\ref{fig:WatchGT}. Unlike Huawei, Garmin shows stress values in real-time. It divides the stress value, in the range of 0-100, into four levels (0-25 as resting, 26-50 as low stress, 51-75 as medium stress, 76-100 as high stress); however, except for resting which is represented as blue, all other levels are  represented as yellow, and are not further distinguished with different colors, as shown in Fig.~\ref{fig:Garmin}. After synchronization, users of both Huawei and Garmin's products can see more stress details with the corresponding mobile applications (Huawei Health, Garmin Connect), including all-day stress data, long-term stress data, the proportion of stress level among other information. % Specially, Garmin shows some activities that may affect stress in theall-day stress graph, such as walking, sleeping, alarm clock, etc. 
These products also provide functions to help users relieve stress, such as deep breathing, biofeedback games, and mindfulness.

\section{Related Work}
\subsection{Stress Sensing and Management}
In HCI, there are many works on innovative approaches to automatic stress-sensing. For instance, office devices such as a mouse and keyboard are explored to detect stress, but this approach is limited to the work environment and cannot be used in other scenarios of daily life \cite{hernandez2014under, vizer2009detecting}. In addition, smartphones embedded with various sensors are commonly studied to detect stress, e.g. based on human voices \cite{lu2012stresssense}, smartphone usage data \cite{vildjiounaite2018unobtrusive}, behavioral metrics of mobile phone activity combined with contextual data \cite{bogomolov2014daily}, and so on. However, they are either too constrained by environmental factors (e.g. can't be too quiet or too noisy)\cite{lu2012stresssense} or too limited in accuracy \cite{vildjiounaite2018unobtrusive} to be used for daily stress detection.

In contrast, wearable devices embedded with sensors are advantageous  for daily stress detection because they can collect objective physiological data and provide relatively timely feedback. For example, electrocardiograph (ECG)  and respiration data obtained from a chest belt \cite{hovsepian2015cstress}, and pulse rate variability features collected from a watch can all been used to detect stress with a satisfactory accuracy rate in the field, and the wrist device is even more portable and less invasive to be used on daily basis.

In addition, much of the work is on biofeedback and intervention technologies to help people relieve stress. Some explore real-time feedback on interventions (such as taking a deep breath) to reflect on their behavior patterns \cite{sanches2010mind}. Others compare the relief effects of different types of interventions including haptic feedback, games, and social networks \cite{paredes2011calmmenow}. Visualizations of stress data with contexts (such as activity and location) are also explored to inform the content and just-in-time interventions \cite{sharmin2015visualization}. However, it is found that the methods of stress feedback need to be carefully designed, otherwise they potentially become stressors \cite{maclean2013moodwings}.

Although not specifically focused on stress,  in work on mental wellness, stress management is an important theme. One focus is on improving mental wellness or peacefulness of mind as a way to deal with stress, e.g through mindfulness \cite{praissman2008mindfulness}, or methods to maintain users' attention \cite{knight2001relaxing,clark1990effects,paredes2011calmmenow}. Some of these studies were at the intersection of mental health and stress tracking, and some were design studies, using focus groups \cite{kelley2017self}, or workshops \cite{lee2018mindnavigator}, and exploring design opportunities for self-tracking.

Overall, the work on stress has mainly focused on innovative approaches to automatic stress-sensing and design that can help relieve stress. Although many stress-tracking products have become commercially available, there have been few studies on the use of these technologies in real life.  Adams et al. conducted a study comparing three stress tracking approaches in the real-world environment (self-report, EDA, and voice-based), and while the study found that these three approaches are about equally effective in different contexts \cite{adams2014towards}, the study didn't evaluate automatic stress tracking. In another instance  a feature analysis of 26 stress management apps investigated how the apps support reflection and action \cite{ptakauskaite2018knowing}; this study also left automatic stress-tracking out from its analysis. As such, there is still a lack of empirical study and understanding of automatic stress-tracking in everyday life. 

\subsection{Data Engagement with Self-tracking Technology}
While little has been done on stress-tracking technology in use in particular, there has been extensive research on the use of other automatic self-tracking technologies, exploring issues of adoption/abandonment \cite{rapp2016personal,epstein2016beyond,clawson2015no, rooksby2014personal}, use in  particular domains such as sports \cite{patel2015contextual} and diabetes management \cite{mamykina2006investigating}, and  design for particular tracking \cite{matthews2015situ, epstein2017examining}. Here we  review the self-tracking work that is related to data engagement. 

Studies of self-tracking technologies in use reveal various challenges to data engagement. These challenges are often explored as barriers to adoption, especially for non-experienced users. For example, Rapp and Cena focused on how novice users perceive and use self-tracking tools in everyday life \cite{rapp2016personal}, revealing a number of data engagement issues that prevented effective integration, including perceived inaccuracy and untrustworthiness based on the users' memories of their behavior, emotional disconnection from abstract visualization, and not knowing what to do with the data due to a lack of suggestions. Similarly, Lazar et al. found that participants abandoned their devices because they did not think the data provided anything informative (e.g. when they went to sleep and when they got up), or they did not know what to do with the data (e.g. what to do with the heart rate data)  \cite{lazar2015we}. Ravichandran et al. 's\cite{ravichandran2017making}  study of sleep tracking technology found that users' misunderstanding of what constitutes  good sleep  restricted them from taking meaningful action. In general, it has been found that, except for quantified selfers who are keen on tracking and numbers \cite{choe2014understanding}, most people find it difficult to engage with  tracked data \cite{rapp2018designing} for  reasons that include incomplete tracking, having too much or too little data \cite{jones2018dealing}, poor aesthetics, unsuitable visualizations, a lack of time, a lack of motivation,  and a lack of related expertise \cite{harrison2015activity,li2010stage,lazar2015we}.  As such, users face many  challenges when leveraging the devices' quantified data for effective use. 

To support data engagement, visualization and integration have been commonly employed to help people gain insights from data, e.g. making the data more ready to use, integrating more contextual information, and correlating different sources of data. For example, Li et al. investigated incorporating contextual information to  the self-tracked performance data, such as steps, to further promote self-awareness and help people find ways to integrate activity into their lives \cite{li2009beyond}. In the study of diabetes self-management practices \cite{mamykina2006investigating}, Mamykina et al. emphasized the importance of a correlation between daily activities (such as exercise, food intake) with the blood sugar levels to help users reflect and make appropriate  lifestyle choices in the future. MONARCA \cite{frost2013supporting} is a system designed for people with bipolar disorder to collect subjective (such as mood, sleep, medicine taken) and objective (such as calls, text messages, physical activity) data through a semi-automatic method that helps them identify factors that may affect their disease. Overall, these approaches primarily  focus on supporting the mental cognitive processes of data engagement by making data more available, visible, and integrated.

%technical implimentation focusing  and and we will mainly review work at theSelf-tracking is approved to be an effective means for life management and improvement in various aspects \cite{lupton2018much, bravata2007using,lin2006fish, ravichandran2017making,epstein2017examining}. In recent years, more tools have been explored to help enhance human capabilities for self-tracking, also called personal informatics [A stage-based model]. For example, while some still require manual input, others have relied on automatic ways for tracking. Quite some work is explored under the rubric of persuasiveness and behavioral change \cite{fritz2014persuasive}. Theories and models for behavioral change has been often drawn on \cite{consolvo2009theory}, and accordingly design features such as goal setting \cite{lin2006fish, munson2012exploring}, rewards for changes \cite{consolvo2006design}, and competition \cite{anderson2007shakra} are commonly explored to increase persuasiveness. Generally, health and exercising are the data commonly tracked.     As such, knowing is mainly explored as instrumental for the purposes of adoption or behavioral change,

In recent years, various novel approaches or designs to support data engagement have also been explored. To address users' challenges, such as low graphical literacy and the inability to uncover subtle correlations between data sets, approaches beyond visualization, including the use of natural language summaries based on statistical analysis, have  been  explored to increase data engagement and understanding \cite{kay2014challenges, bentley2013health}. It has been suggested that some features, like dialogue, might influence the need for reflection \cite{halttu2017persuading}. Social approaches to reflection have also been investigated. For instance, Feustel et al. looked at the idea of aggregating cohort data into personal informatics systems to support meaningful reflection \cite{feustel2018people}, and Graham et al. conducted a study to understand shared reflection by asking people to reflect on each other's data \cite{graham2016help}. 

Since the issues surrounding the data engagement of self-tracking is often explored in terms of reflection \cite{li2010stage}, the conceptual meaning of reflection has  also systematically been reviewed and examined to support the design for reflection \cite{slovak2017reflective, baumer2015reflective, ploderer2014social}. For instance, drawing on Schon's notion of reflective practum, Slovak et al. identified three components to scaffold for reflection, explicit (the link between experience and reflection), social and personal components \cite{slovak2017reflective}. Baumer reviewed conceptual and theoretical models of reflection, and identified three dimensions: breakdown, inquiry and transformation \cite{baumer2015reflective}. Ploderer et al. distinguished between reflection-in-action and reflection-on-action, with the former referring to the reflection  of realtime feedback, and the latter to data exploration when convenient \cite{ploderer2014social}. However, while insightful, these meanings have evolved from general theories or empirical studies of reflection, and do not involve self-tracking technologies, and thus may miss the unique complexities and dynamics of reflection brought by the involvement of the self-tracking technologies themselves. 

In this paper, we focus  on a complex and relatively new and under-explored automatic tracking technology -- stress tracking  -- in use, and hope to uncover new insights into users' data engagement with self-tracking technologies in practice to contribute to this body of work and inform the related design of self-tracking technologies.

\section{Methods}

To gain a better understanding of the use of automatic stress-tracking technologies in practice, we adopted the qualitative research method to uncover the rich and detailed usage data, by interviewing those who had already used and experienced related technologies. For participant recruitment, we designed a flyer searching for those who had used wearable devices with stress-tracking features. On the flyer, we described our study motivation, study method, participant qualifications and compensation, and one of the authors' WeChat contact information. In China, the two most popular brands of smart wearable devices with stress-tracking features are Huawei(including Honor) and Garmin \cite{topwatch, topbracelet}, so we posted the recruitment flyers in the QQ and WeChat groups for Huawei and Garmin wearable device users, as well as some general wearable smart device communities, and our own WeChat circles to recruit more users. Finally, we recruited 17 participants in total, including 14 from the user groups or communities, and 3 from participants' recommendations. Their profile information is shown in Table 1. Most of them are male, and include 16 males and 1 female, largely aligning with the male to female ratio of smart wearable device market \cite{tencent2015wearable}. In addition, most are young, ranging from 21 to 39 years old. Their occupations are diverse, and include those related to science and technology, such as IT manufacturing trainee, technical developer, programmer, wearable health worker, and exercise physiology worker, as well as those that are not IT related, including salesman, government worker, and customer service employee. As shown in the table, they were from many different cities in China, ranging from inland cities in the north, such as Beijing, Zhengzhou, Shenyang, and Xi'an, as well as coastal cities in the south, such as Dongguan, Shenzhen and Guangzhou. Most of the participants used Huawei's Watch GT or Honor Watch Magic; P12 and P17 used Garmin's Vívosmart 4 and  Forerunner 645 Music, respectively. All had used the devices from between half a month to 12 months. 

%\hl{They live in many different cities. Beijing, Zhengzhou, Xuchang, Shenyang and Xi'an are inland cities in the north of China, while Shanghai, Dongguan, Shenzhen and Guangzhou are coastal cities in the south of China. Among them, Beijing, Shanghai, Shenzhen and Guangzhou are the most economically developed cities in China.}

\begin{table*}
  \caption{Participants}
  \label{tab:participants}
  \begin{threeparttable}
  \begin{tabular}{cccccl}
    \toprule
    ID&Gender&Age&Occupation&Device&Location\\
    \midrule
    P1 & M & 21 & Exhibition Salesman & Watch GT & Shanghai \\
    P2 & M & 31 & Government Worker & Honor Watch Magic & Tonghua\tnote{*} \\
    P3 & M & 21 & IT Manufacturing Trainee & Honor Watch Magic, Watch GT & Dongguan\tnote{**} \\
    P4 & M & 35 & Technical Developer & Huawei B5 bracelet & Beijing \\
    P5 & M & 22 & Design and Operation Worker & Honor Watch Magic & Zhengzhou\tnote{*} \\
    P6 & M & 28 & Programmer & Honor Watch Magic & Beijing \\
    P7 & F & 26 & Wearable Health Worker & Watch GT & Shenzhen \\
    P8 & M & 31 & Business Operator & Honor Watch Magic & Xuchang \tnote{*}\\
    P9 & M & 39 & Government Worker & Watch GT & Shanghai \\
    P10 & M & 26 & Programmer & Honor Watch Magic & Beijing \\
    P11 & M & 28 & Unemployed & Watch GT & Shenyang\tnote{*} \\
    P12 & M & 25 & Exercise physiology worker & Watch GT, Garmin vívosmart4 & Finland and China \\
    P13 & M & 34 & Customer service & Watch GT & Shanghai \\
    P14 & M & 32 & Unemployed & Honor Watch Magic & Shanghai \\
    P15 & M & 28 & Programmer & Honor Watch Magic, Watch GT & Xi'an\tnote{*} \\
    P16 & M & 24 & Advertising designer & Watch GT & Shanghai \\
    P17 & M & 24 & Student & Garmin Forerunner 645 Music & Guangzhou \\
    \bottomrule
 \end{tabular}
     \begin{tablenotes}    
        \footnotesize         
        % \item[*] coastal cities in southern China, others are inland cities in northern China, except for P12's location.
        \item[*] inland cities in northern China
        \item[**] coastal cities in southern China
     \end{tablenotes}          
 \end{threeparttable}
\end{table*}

We then conducted semi-structured interviews with these participants. Since the participants lived in different cities, most interviews were through WeChat voice calls, except for P9, P13, and P14 with whom we did interviews face-to-face. The interviews usually lasted about 40 minutes. During the interviews, we asked participants for basic information including their age, occupation, location, and education, as well as questions of their general use of devices and how they experience and manage stress, such as what devices they had used, what applications in the devices they used most frequently, how  they used these applications, their stress status, their perceived stress source(s), and how they deal with stress on a daily basis. We then asked about details of their use of stress-tracking technology, probing for concrete usage instances, e.g under what circumstances they checked the stress data and how, what they saw and how they experienced and understood it, what they did after seeing the data, etc. For some usage instances, we asked whether they could provide screen shots of their stress application interfaces for clarification, and some  sent screen shots over WeChat to us. We followed up with some of the participants after the interviews, keeping in touch  through WeChat to know more about their use of the devices,  special events they found in their later use, and  changes in their long-term stress status. We also collected these chats for later data analysis.

All interviews were conducted in Chinese Mandarin. With the consent of the participants, we audio-recorded the interview process and transcribed it into text for later data analysis. For privacy purposes, we anonymized their data in the transcript and in the paper.

%\subsection{Reviews Data on Amazon}
 %In order to corroborate our findings, we also collected the review data on stress tracking technology on Amazon, to go beyond the Chinese context and the limited devices we cover in the study. We wrote a crawler program and crawled the review data of products that we are aware with stress tracking features including Garmin, Samsung, Gobe2, Wellbe, Spire Stone and Bellabeat Leaf Urban on Amazon and filter out reviews that do not contain ``stress'' when crawling. Finally, we collected 982 reviews and saved them to a CSV file for later analysis, including 152 for Garmin, 39 for Samsung, 29 for Gobe2, 4 for wellbe, 525 for Spire Stone, and 233 for Bellabeat Leaf Urban. The average number of words in these comments is about 170.

We conducted thematic analysis inductively \cite{braun2019reflecting} with the interview data. We first familiarized ourselves individually with the data and then extensively read, analyzed, and discussed it together. Each of us generated our own set of codes, and we compared our codes in meetings and discussed it further. We eventually identified the challenges of engaging and understanding data as the primary theme. We identified three sub-themes under this key theme, which we report in our findings. For privacy purposes, we anonymized our interview in the paper by using P\# to represent the interview participants.

\section{Findings}
%围绕stress tracking的特殊性来写，主管感受，subjective， 与physical 特征，

Almost all of our interview participants had already adopted and integrated their smart wearable devices into their everyday lives, so we do not have  adoption issues as discussed in previous works (e.g.\cite{rapp2016personal}). Our participants wore their watch or bracelet all the time except for occasions when it was not feasible, such as when it was charging or they were taking a shower. All of our participants, except P2 and P10, had not bought the watch or bracelet for the primary purpose of stress-tracking, but for other reasons including  sports, to not miss phone calls, or to replace a traditional watch. In  fact, most  were not aware of the existence of the stress-tracking feature at the time of purchase and only discovered it later while exploring the device. P17, who loved jogging, offered a typical explanation: ``\textit{I didn't even know there was a stress-tracking feature. I didn't buy the Garmin watch for stress detection. I bought and used [the watch]. It was after I used it that I got to know this feature.}'' Although participants did not adopt the technology for the sake of tracking stress, they all quickly became aware of the feature as it is quite accessible, just a few clicks or swipes away. 

%who was concerned with long term stress or P10 who was curious about the stress tracking feature,
%P10 was such an case. He was very excited and curious about the stress sensing feature, which had become part of the reason for him to purchase the smart watch: ``\textit{I've never had (other stress tracking technologies), and this is the first one. I bought a watch partly because I saw its stress function. It is very novel that stress can be measured. It is very powerful.}''
We also found that, while participants could easily access the stress data, their understanding of it was quite varied. Only a few could meaningfully engage with it, some only had a limited understanding, while others were confused. At the same time, several participants reported that just the awareness of the existence of the stress tracking feature, not necessarily an understanding of the data, had some impact on their behavior, as similar to what is found in \cite{fritz2014persuasive}. For P13,  simply being aware that the watch constantly monitored his stress helped him to watch his temper: ``\textit{So at that time when I did not have the watch, I would not deliberately control my mood or emotions, and I would just discharge. After wearing this watch, to some extent, I felt that I was monitored every day, so I couldn't make myself too stressed, or lose my temper.}'' Similarly, to P10, the awareness of the feature had impact on his behavior subconsciously: ``\textit{If you have [the stress tracking feature], you will subconsciously adjust yourself...}'' However, not being able to effectively engage with the stress data overall kept most of them from making more informed use of it. Below, we turn our focus to the challenges of stress-tracking data engagement we uncovered from the study.

\subsection{Lack of Immediate Awareness}
All of our participants were excited about the automatic stress-tracking feature when they began using their devices. They checked the data frequently. However, after one or two months, the novelty effect was gone, and the majority (except P12) stopped engaging with the  data in a timely and frequent manner. P2's situation is typical among our participants: ``\textit{I paid close attention to the stress data when I just started wearing the watch. I checked the data at least ten times a day, but I checked it less and less frequently over time. It's been 2 months since I started using it, and now I only check the data twice per day.}''

The main reason for this was that our participants did not feel that the stress-tracking devices were helpful to raising their immediate awareness due to their natural responses to some stressful events and the limitations of the devices. First, when people encounter challenges, their mind is usually too occupied by distressing thoughts \cite{elkin2013stress} to break away from their minds and check the stress-tracking data. Participants (except P12) commonly reported that when they were facing challenging issues, they were overwhelmed and would intuitively focus on solving their issues rather than checking the stress data. For instance, P3 told of a time when he got assigned a challenging task at work that made him feel so stressed that instead of checking his stress level, he focused on the task first: ``\textit{I was assigned with some tasks last time, when I just started my internship. wow, what the hell! I could not believe it. I then asked others a lot of questions and looked up information. I was very stressed ... But when you are truly stressed, surely you don't think of looking at the watch  – you think of solving the current problem first.}'' Similarly, P5 noted: ``\textit{When I am busy and stressed, to be honest, I don't pay much attention to [the data]. I would probably just check the time on the watch at best.}'' 

In rare cases, when participants were  stressed to the point of physical discomfort, however, they might look at the data in that moment. As P2 reported: ``\textit{When I felt that my heart was beating a bit fast, I would take a look at it. Or sometimes, when I was writing a lot of summaries or textual materials, and felt dizzy, I would take a look at it and pay more attention to stress.}'' However, in these cases,  devices that don't provide real-time updates made it difficult to gain an immediate awareness. For example, Huawei's stress-tracking app only updates every half hour, and P12 reported how it caused confusion sometimes as the data was not consistent with what they felt in the moment, ``\textit{You can feel  your own heartbeat rising. You call a customer, or do an interview, and you feel your heartbeat rising, but when you look at your watch, the stress data has not risen.}'' Our participants complained how Huawei's delay discouraged them from checking it in the moment they felt their emotions intensify.  P3 said: ``\textit{The problem is... I think it should update more frequently... If your stress rises again within half an hour, it can't be monitored at all...}'' As such, this lack of timely feedback makes it even more difficult for people to develop an immediate awareness of their stress  status.

%\hl{P12 had used Huawei (non real-time stress sensing) and Garmin (real-time stress sensing) watches, and he worked in Finland. The use experience of different stress tracking technologies, working environment and culture may lead him to timely check his stress data, which is different from most of our participants.} 

Considering it is challenging for people to remember to use their stress-tracking devices and the importance of in-the-moment interpretation, it is critical  that stress-tracking devices help raise immediate awareness. However, our participants reported that the devices failed to notify them when their stress levels were high or changed remarkably. This made our participants feel like the devices were not helpful. For example, P4 explained, 
\begin{quote}
    ``\textit{I now feel this function is not so meaningful…since it doesn't react or intervene in time. I often only realize that I was stressed out after my stress has gone…Such devices should help people manage real-time stress rather than just recording it, right?...Only recording it is not so helpful. I was interested in checking the data in the beginning because it seemed to be a novel function, but now, since it's not so helpful, I don't check it frequently anymore.}''
\end{quote}
Even P12, who was the most engaged user among all of our participants, said, 
\begin{quote}
    ``\textit{Both devices (Huawei Watch Gt and Garmin watch) claimed they have something like a stress level reminder, but I have never been reminded...even when my stress level was as high as between 80\% to 90\% in Garmin...I wish they could provide an in-the-moment reminder and provide us effective, timely ways for stress management, such as providing relaxing music for stress relief...If they could remind us that our stress was high in the moment, we would have more interactions with the data and could manage our stress better…}''
\end{quote}
 Users' in-the-moment data engagement is critical for reflection \cite{slovak2017reflective} and intervention. Failing to provide timely reminders hinders users from effectively managing their stress.

Most of the time, they noticed the stress data through a casual or random encounter with the technology. That is, they did not intentionally check the data; rather, they only noticed it when they were browsing  other types of data (e.g., heart rates) on their devices, when  casually playing with their devices when  bored, or when  just taking a glance  at the devices while taking them off. During these casual encounters with the  data, something would stand out and become noticeable, drawing their attention to it. Most commonly this was a  sudden rise in stress level or a change of color. For instance, P10 once noticed an unexpected, sudden rise in his stress level after lunch after having a period of relative stability: ``\textit{ I usually take a nap during the lunch break, and usually my stress is quite stable. However, there was one time, after having a nap, I saw a sudden rise of stress level when I was randomly playing with my watch. The stress was quite high.}'' P5's stress data caught his attention due to its change of color: ``\textit{ Yeah, on the 13th of this month, just two days ago…I saw the yellow color for the first time. The colors range from blue to green to yellow to red. Yellow means the stress is quite high. It was the first time for me to see such a high value.}'' As shown in these cases, the visual presentation of the data (e.g. the bright color over a dark background, the sudden change) against the participants' personal experiences (the first time seeing it) led them notice it. As a result, while the automatic stress-tracking devices afford rich data, only a very small portion of it actually drew our participants' attention and motivated them to interpret or reflect on it. 

As these cases also illustrate, not all quantified numbers receive the same attention, or have equal importance – only some of the information stands out and  matters. Data presentation or visual design plays a certain role here in filtering out this information and telling users where to draw their attention. However, the participants' attention was usually only drawn to the data during casual encounters. Overall, our participants did not engage with the data frequently  because the devices failed to raise their immediate awareness and provide effective in-the-moment interventions. 

\subsection{Lack of Pre-required Knowledge}
When some of the data actually drew our participants' attention and motivated  them to interpret or reflect on it, most of our participants (except P12) found it challenging to make meaning out of it. This is primarily because our participants had adopted the psychological notion of stress, while the devices measured physiological stress. In other words, our participants perceived stress as a subjective concept that refers to the feeling and perception of pressure while the automatic stress-tracking devices measured stressed physiologically by analyzing  bodily reactions generated by the autonomic nervous system. Unlike tracked activities, which are more straightforward for quantification and interpretation, such as steps and hours of sleep, measures such as stress are often more challenging for interpretation. The distinction between the psychological notion of stress that our participants adopted and the physiological notion of stress that the devices were based on led to barriers that prevented  our participants from interpreting and making meaningful use of the tracked stress data.

Among our participants, only P12 understood that the devices were based on HRV and measured stress physiologically rather than psychologically. He managed to learn the related technical and domain knowledge early on by searching online and reading related scientific articles: ``\textit{I read some articles on the Internet, and  some popular science articles about what the autonomic nervous system is and the relationship between the heart rate variability and the autonomic nervous system.}'' So he had the basic understanding that the stress measured by the devices corresponded to how  his body responded to external demands. However, our other participants were confused when interpreting the data, since it didn't reflect their subjective feelings of stress, that is, their perceptions of pressure. 

In everyday conversation, when people say ``stress,'' it usually means psychological stress. Asking to fill out a stress questionnaire to use this feature on some products such as Huawei's further reinforced such a perception. Thus, our participants (except P12) felt that the device wasn't helpful  when they discovered that the measurements from the tracking technology did not match their feelings. When triggered by ``feeling something'', our participants often expected to see  that reflected on the tracking technology  and became disappointed when it wasn't. P11 explained: ``\textit{I had just bought it, and at that period of time I was under great pressure; however, it did not show it when I felt stressed several times. I can't remember what happened exactly. I just remember that I specifically looked at it when I'd just bought it and felt stressed but it didn't change as much as I'd imagine.}'' P13 had a similar experience and thought that the stress-tracking application was inaccurate: ``\textit{I found it was inaccurate when I began to use it. I was unhappy and lost my temper at that time, and I found (my measured stress was) just 'medium' instead of 'high'.}''

Moreover, some participants found that there was often a correlation between their physical activities, such as eating and exercises, and changes in their stress levels on the device, which baffled them. P1 noticed that his detected stress level rose after lunch: ``\textit{I don't think it's accurate...Most time it is around noon, such as after eating, the stress is higher. I don't understand why... }'' Only after we explained that the device was based on HRV and that eating could put a physiological burden on the body because of  digestion, did he think it made sense. Similarly, P6 thought that a lot of things in his life stressed him out and was puzzled why he did not see them being manifested in the application. He wondered what counted as stress:
\begin{quote}
    ``\textit{How could there be no stress in my life? I need to pay a mortgage monthly, which is definitely stressful. I'm quite worried every time I think about it. [Yet] This isn't reflected [on the watch]. Doesn't the situation count as stress? I don't understand. It is not reflected anyway. If it could be reflected, I think the watch would show my stress level as yellow everyday.}''
\end{quote}
As such,  the complexity of stress -- e.g. involving both psychological and physiological, the external and the internal -- makes interpreting its data more difficult than other tracked data, such as steps and calories, on the same device. 

Oftentimes, the displayed stress range (e.g. ``relax'', ``normal'', ``medium stress'' or ``high stress'') did not fall into our participants' subjectively felt understanding. For example, P2, whose stress level had been high on the application, was doubtful as to its accuracy, as the application never displayed a  ``low'' level, even when he was engaging in relaxing activities: ``\textit{…for instance, when I go out to watch a movie, eat, or have fun, like sing karaoke with friends, of course I don't have stress, but the watch still showed that my stress was high....}'' P2 even tried to recalibrate the measurement by retaking the stress questionnaire:
\begin{quote}
    ``\textit{I always doubted that whether that's because my answers to the questionnaire were too pessimistic, and the watch thus set the baseline stress scale higher than how I truly felt. Thus, I unbonded it with my account…and retook the questionnaire. I intentionally answered the questions more positively than the first time. It turned out that the measured stress levels overall have indeed decreased a bit, but are still higher than how I feel. I'm confused. Why is it always high,  whether I am relaxed or not?}''
\end{quote}
When evaluating whether the stress data was accurate or not, P2 was comparing the data against how he felt, i.e, his  psychological stress level. The mismatch between the physiological type of stress that the device measured and the psychological stress that P2 perceived led to P2's confusion. Similar to what is found in previous work \cite{epstein2016beyond, lazar2015we}, this kind of  perceived inconsistency caused our participants to distrust the system and even led some to stop paying attention to it. 

To make things worse, the underlying technological mechanism of the application was not so straightforward either. In our interview study, many expressed that they did not know how the stress was sensed by the devices. They had different kinds of speculations. For instance, P1 asked us, ``\textit{Is it based on some kind of algorithms to calculate my stress level? Or is it monitoring my blood pressures or something through my skin?}'' Some assumed it corresponded to their real-time heart rate.  P7's interpretation was typical: ``\textit{My understanding is that it mainly depends on your heart rate...For example, if your heart rate is relatively high, it will recognize that you may be a bit more stressed.}'' P3 also guessed that the measured stress level was determined by the heart rate, noting ``\textit{If the stress is higher, the heartbeat will be faster.}'' However, this theory soon led to further confusion, as participants  discovered it was not exactly right: ``\textit{In the afternoon I went to other places and took a look. I had a lot of activities, but the stress was not high. I don't think the stress is based on the heart rate.}'' P5 went through a similar process when he realized:\textit{`` My heart rate was high during exercise, [but] the stress value was normal.}'' Almost all our participants wanted to learn more about the underlying mechanism behind the stress-tracking. For instance, P17, when asked whether there was anything he did not understand, explicitly told us that he had agreed to be interviewed because he wanted to find out how the application detected stress: ``\textit{How is it measured? It should be calculated by some algorithm, but I don't know what specific algorithm it is... I'm definitely curious. That is why I accepted [your interview], because I'm curious.}'' 

Additionally, the conditions for stress-tracking also caused more confusion, as the devices only sensed stress when one was still. P12 reported such confusion: ``\textit{In the beginning, I didn't know why my stress value was not shown at noon. It was weird. You see, the stress value usually disappeared from 11:00 to 12:00, and it came out again from 12:00 to 13:00. It was strange. Why did it usually disappear at 11 o'clock?}'' Only after we explained to him, did he realize it was because he was actively moving around noon that stress data was not detected. In a word, the not so intuitive mechanism for stress tracking led more confusions. 

In summary, even when our participants paid attention to and tried to interpret the data, it was challenging for them to decipher it due to the mismatch between what the devices measured and what our participants considered to be stress, as well as the unclear underlying technological mechanisms of the automatic stress tracking.

\subsection{Lack of Communal Support} 
Despite that our participants encountered challenges when trying to make sense of the stress data, they did not have easy access to related resources and communal support to help them tackle the challenges to achieve meaningful interpretations.  

Alone among our participants, P12, who lives in both China and Finland, was primarily working in Finland at the time, represents a contrasting case. When working in Finland, he was situated in a social world which helped him develop a shared understanding. He described, ``\textit{In our company, quite some employees are wearing sports bracelets. I feel 60 percent are wearing these, so it is also part of our topic, and we chat quite much about it...[stress tracking] is a function of our lives and is something we all use.}'' The socialization at P12's company also helped him to understand the application better. He explained, ``\textit{I had a cup of coffee when getting up in the morning and I felt relaxed, but my watch showed my stress was medium. So I asked my colleagues, `Is it because of the coffee?' and my colleagues said it was...}'' Without understanding that stress is not simply a psychological concept but also a physiological one, it might not be so easy to see that drinking coffee can cause stress levels to rise; this would be even more difficult for  someone to process if  they actually felt relaxed after doing so. More studies have shown the importance of socialization or social processes for learning and forming shared background understanding for interpretation (e.g. \cite{lave1991situated}). P12's experience is a case in point. Being part of such a community provides the social means to acquire related knowledge and collectively address matters of confusion.

Unfortunately, for the other participants who were all in China, corresponding stress-oriented communities of practice still have not yet formed  to help develop the need for shared understanding. In China, while sports-related tracking technologies have become quite popular, and many social groups have formed, stress-tracking is still new and something of which people are rarely aware. However, P17 provides a nice example that illustrates how being part of a community of practice can make a difference.  P17, who joined several sports-related groups and bought the watch for sports as most of members in the group did, reported how he could easily interpret and meaningfully read the tracked numbers for pace and heart rate:
\begin{quote}
    ``\textit{I just check to see if my pace matches my heart rate. If you know how to run, then (you will know), for example, if I run at a pace of 6, then my heart rate should be about 140. If I run at a pace of 6 one day but my heart rate suddenly reaches 150, I will know that my athletic ability has dropped; if my pace is 6, and my heart rate becomes lower, e.g. it was lower than the previous 140 and was 130, I will know that I have improved.}''
\end{quote}
By looking his pace and heart rate, he could easily tell whether his athletic ability had improved or not. On the other hand, as he explained, it was not easy for him to interpret the stress data: ``\textit{For those who bought the watch just for running, they would not understand it at all. He may only understand that the Chinese words or number there shows something about stress, but he wouldn't know what exactly the words or the number means. There is no way for me to know, and I believe most people wouldn't know either. I only have a vague knowledge about it.}'' As such, while he could meaningfully read the running- related numbers, the stress numbers  still  puzzled  him. 

In other words, the broader social and cultural context shapes how people approach the tracked stress data and whether they understand it. P12 described the different accessibility of related learning resources about sports and about stress in China today: ``\textit{For example, there are many books about running on the Internet in China, but there have not been books really about life stress...In China, I think there is still a lack of knowledge about stress management or life management.}'' P12's observation was confirmed by our other participants' experiences. P2, who adopted the watch to learn more about his stress, only knew that the watch confirmed that his stress level was generally high, but  did not know how to deal with it. Some participants reported that they only wanted  to know whether their stress level was normal or not, but they were usually unable to tell by just looking at the data. This is to say, due to the lack of support from the broader socio-cultural context in China, it is not easy for people to develop a meaningful reading of the data beyond whether or not their stress level is high or low. 

\section{Discussions and Implications}

% the lack of awareness is made more salient with the comparison of two types of devices, the lack of real time feedback
%女生少，只有一个，男生有解释一下，设计有关，外观比较偏运动，

In the preceding sections, we presented a study on the use of the automatic stress tracking technology in practice, highlighting three challenges presented by its data engagement: a lack of immediate awareness preventing engagement with data in-the-moment, a lack of pre-required knowledge, domain and technical, and a lack of communal support. As shown here, these challenges are not merely associated with one's capabilities, such as graphical literacy or quantitative analysis capabilities as pointed out in prior works \cite{rapp2016personal}, but have to do with factors embedded in corresponding social practices of stress tracking with the technology, such as people too occupied to check the stress data in the moment when they were stressed, the technology failing to provide timely feedback and reminders, as well as the mismatch between the scientific notion of stress and the everyday use of stress. While previous works revealed similar challenges, such as a lack of the expertise needed to interpret the tracking data \cite{lazar2015we}, these challenges were mainly identified as reasons for adoption or abandonment. Data engagement itself, and its association with corresponding social practices, have not received sufficient attention. Drawing on our study of the use of the automatic stress tracking technology in particular -- a relatively recent development and more complicated technology, we made data engagement our focal point, and unpacked the underlying reasons contributing to these challenges.

\subsection{Casual Encounter Mode of Data}
As shown in our study, one challenge of stress-tracking data engagement comes from \emph{how the data is encountered}. Automatic tracking has been primarily  explored to relieve the burden of manual tracking \cite{choe2017semi}. However, just as Choe  \cite{choe2014understanding} and others (e.g. \cite{li2012using}) have pointed out, automatic tracking can reduce engagement and awareness, which can also compromise its effects. More importantly, as revealed in our study, the users' intention to track the data, that is often assumed in self-tracking work, is simply not there anymore for the stress tracking feature coming with the wearable device. But rather, they noticed their stress data while engaged in their daily non-stress-oriented practices. This is quite different from those who deliberately engage in an active management practice, with the tracking technology intentionally employed to facilitate a process, e.g. to help address their insomnia issues \cite{ravichandran2017making} or manage a chronic illness such as diabetes \cite{mamykina2006investigating}. We can label this  \emph{casual encounter mode }to distinguish it from serious encounter mode in which users have the intention to track and engage with data.

When the users' intention to track data can simply be assumed, we can reduce self-tracking data engagement to a mere analytic issue, as has often been the approach in prior work for data engagement \cite{li2009beyond, mamykina2006investigating, frost2013supporting}. In the casual encounter mode, however, users just encounter the data being collected and presented to them. Instead of pulling the data out for analysis themselves, users are pulled towards certain data that draws their attention. As a result, only a small amount of data will become ``present'' to them, and what data becomes present is highly dependent on how it is presented and the interactive practices users use to engage with the devices. That is, these two different modes, casual encounter and serious encounter, are totally different in terms how and what data will come to the users' conscious attention for engagement. With more and more automatic self-tracking technologies embedded in  everyday objects such as smart watches, we believe the non-intentional and casual encounter will become increasingly more commonplace. To address the data engagement issues in casual mode then, we should go beyond simply user interface revisions which would not make these devices work, and consider the practice as a whole. Below we discuss implications based on the casual encounter mode and the challenges we identified from the study. 

\subsection{Supporting Situated Interpretation}
To support casual users engaging with the automatically collected data, there needs to be not simply new ways of data presentation or integration, but new interactive designs to help engage users in the right moment. For example, in addition to increasing tracking frequency, we, like many of our participants, think  it would be valuable to provide reminder functions at appropriate times, e.g. when there is a big jump or drop in stress level, or when the stress level exceeds a certain threshold, in order to fully support the situated interpretation process. The reminder functions could help users capture, engage with, and interpret the notable data in the moment within the situated context, not from  hindsight. Of course, the reminder should be provided in a peripheral and non-intrusive way, such as a vibration. Design studies should be conducted to identify what moments will be good for sending alerts and how. Also, users should be allowed to customize whether to turn on the alert and under what circumstances need the alert. What we would like to emphasize here is that this is different from integrating contextual information into the data presentation as is often approached, but is an approach that supports situated data engagement and ``reflection-in-action'' \cite{slovak2017reflective} through which one's experience could be directly drawn on for data interpretation and correlations.

%With these challenges, despite the rich stress tracking data made available, most data simply appears  ``transparent'',  yet a small portion could actually become ``present'' to users \cite{dreyfus1991being}, of which an even smaller portion could be meaningfully interpreted.

%Although our study is based on automatic stress tracking in particular, the challenges involved as well as the work and conditions for meaningful data engagement uncovered from our study could have broader implications for self-tracking technology. 
%As shown here, there are three types of understanding: unreflective or transparent coping before \textbf{affectedness}, \textbf{interpretation} in a deliberate but still context and \textbf{forestructure} dependent way after affected, and confusion or deciphering when they encounter something without corresponding \textbf{forestructure}. The three types of understanding can be illuminated this way:\textit''{If one has the linguistic abilities of an average American plus a smattering of German, then one understands English (i.e., can use it transparently), one can interpret German (i.e., use it in a deliberate but still context dependent way), but one must decipher Japanese (i.e., treat it as a meaningless code)}''\cite{dreyfus1991being}(p.198).
\subsection{Making Pre-required Knowledge More Available}
Another challenge comes from the kind of expertise needed for data engagement. As shown in the study, the meaningful interpretation of the stress data requires necessary knowledge, including the domain knowledge of stress and the technical knowledge of the mechanism of stress monitoring.  Intentional tracking and an active management practice often mean that users already have acquired the requisite  knowledge needed for the use of the self-tracking technology, e.g. knowing what glucose or blood sugar level means and how they should be managed. However, with more data being automatically tracked and readily available with the wearable devices, as in our case, this prerequisite can no longer be assumed. Those who sought out the relevant resources to gain related expertise on the Internet, such as popular science articles on stress or videos related to how to use the Garmin watch, were able to overcome some of the challenges. However, for most users, despite their interest and curiosity, there was no easy access to learning resources.

Besides automatic tracking, this challenge also becomes more salient with the nature of stress itself, a physiological state that is more complex and less straightforward than other measures, such as steps. A close examination of our data suggests that sometimes the users' perceived inaccuracy was often due to a mismatch between their subjective experiences and the qualitative presentation of the stress data. For example, when they experienced stress, users expected to see their tracked stress presentation to be ``high'' and match their subjective feelings, rather than ``medium'' or ``low,'' which made them perceive the technology as inaccurate; although the trend of the change in quantitative terms, the rising or dropping of the curve, actually corresponded well with their change of feelings. As such, dealing with  health or physiological states similar to stress, things that can only be experienced subjectively, could cause more trust issues and add more challenges to data engagement. A meaningful reading of this kind of health data, such as stress, heart rate and sleep, thus requires more specialized knowledge and can pose more challenges for lay people. However, as the majority of the work of automatic self-tracking technologies thus far has focused on relatively more straightforward data, these  challenges  have not been sufficiently emphasized. 

Our study also revealed that different layers of understanding can be achieved through different levels of acquired knowledge; while some simply got a clearer idea of their stress level, others, such as P12, developed a more meaningful understanding of how stress  correlated to different aspects of their lives. We can call the former ``direct understanding'' and the latter ``deepened understanding''. The difference between the two is similar to the difference between a lay person and an experienced doctor who can read a lab test result. While the lay person can only tell whether the results are normal or not (i.e., within normative range), the doctor can tell whether the patient's condition has improved and whether the immune system has become stronger. To achieve a ``deepened understanding'', a deliberate effort to acquire related knowledge and expertise is needed. 

Therefore, for the automatically tracked health data, it is crucial to make the requisite knowledge more available and present it in a meaningful way to help users interpret the data. In the case of stress-tracking, it means making it more possible for users to learn more knowledge about stress, more about what stress means in the device, and more about what the mechanism for tracking is, among other things. For a more meaningful engagement and deepened understanding, design that helps people acquire related expertise becomes even more important; we could consider integrating learning materials, in terms of short texts, pictures, or videos, into the products to make them more accessible and to present them in a more compelling way. More work is needed here to specifically understand  how  knowledge could be presented for users to interpret their data more effectively, e.g. integrating relevant knowledge through visualization, leveraging intelligent conversation agents to support user inquiry about the data, etc.

%To make these learning resources more accessible, we may consider integrating these learning materials in terms of short texts, pictures or videos, in the products and present them in a compelling way. For example, when users turn on the device, they will be shown related tutorials in a piecemeal fashion, one piece at a time, such as the hazards caused by long-term high stress, what indicators the stress detection is based on, what is HRV (if based on HRV), and what factors will affect stress detection. In addition, for those who want to know more, we can provide them with information or links to related resources, such as a book about stress, a paper about the principle of stress detection, and the URLs of related popular science videos on YouTube.

\subsection{Supporting Knowing with Communities of Practice}
For meaningful data engagement, it is also important to support the social processes of "knowing." What we have seen in the study is that the concept of stress in science still has not been merged with the concept of stress in everyday life \cite{tudge1992vygotsky}, and methods for stress-tracking and stress management have still not become part of popular culture, adding to the difficulty of interpreting tracked stress data. P7's case is the telling one. While he could easily interpret the sports data and gained a corresponding  understanding from the sports community of practice, the stress data was still puzzling to him. Contemporary anthropological and sociological theorizing has already illustrated that participation in the social world is a fundamental form of human learning/knowing \cite{lave1991situated}. Lave and Wenger, for instance, focus on social engagement and participation as the context in which learning occurs  \cite{lave1991situated}, and call the broader context, or the social world, ``communities of practice''. It is through communities of practice that resources are shared, information spreads, and shared understanding is achieved. Where self-tracking technologies are concerned, we believe participation in corresponding communities of practice  is the key to go from simple tracking to "knowing." 

Recognizing the social nature of learning/knowing provides a different perspective for design, e.g. facilitating the forming of corresponding communities of practice is as  important as making learning resources easily accessible. Think about the recent development of open science \cite{vicente2018open} or citizen participation in the scientific inquiry processes; this provides a valuable model not simply for scientific discovery, but also for scientific education \cite{bonney2016can}, or the merging of scientific and everyday knowing. This is different from social discussions for data analysis \cite{feustel2018people, graham2016help}, and is more of a community that helps members learn through their participation and social interactions. Supporting the formation of communities of practices and user participation, could be an effective approach to the support of learning and "knowing" of self-tracking technologies. 

\section{Limitations and future work}
We note that although the gender ratio of our participants largely aligns with the gender ratio of smart wearable device users in China \cite{tencent2015wearable}, there is only one female participant in our study, which might have introduced gender bias. For future work, with automatic stress-tracking features becoming more available in  wearable devices, it would be helpful to have more studies of this kind look into detailed usage across different sites and diverse populations. Our study also suggests promising directions for future design explorations, mainly to address the data engagement challenges identified in this work, e.g. mechanisms to support in-situ data interpretation, effective ways to integrate corresponding expertise knowledge with the data, and ways to develop communal support to help users form a shared understanding.

%\For future work, with automatic stress-tracking features becoming more available in our wearable devices, it will be helpful to have more studies of this kind to look into the detailed usage across different sites, to help examine the resonances among them.

\section{Conclusions}
 In this paper, we present a study of the use of a relatively recent and less straightforward stress-tracking technology-in-practice, highlighting the three primary challenges of data engagement with automatically-tracked stress data: a lack of immediate awareness, a lack of prerequisite expertise, and  a lack of corresponding communities of practice. In particular, by focusing on a relatively ``unfamiliar'' stress-tracking technology, the study reveals that many assumptions that have been taken for granted about self-tracking technologies no longer hold true in the rapidly changing world. Reflecting on the challenges uncovered from our study as well as related works, it is clear that some elements of the technically-mediated tracking practices make a difference as far as data engagement is concerned. These include how the data is tracked /encountered -- intentionally or unintentionally; who the users are -- novice or expert; what is track -- activity or health states; and where the users are situated -- in a community of practice of corresponding tracking or not.
 
 %In recent years, self-tracking technology has become increasingly ubiquitous, and has attracted extensive research attention. Data engagement in self-tracking has been mainly approached with the notion of ``reflection''. While ``reflection'' is central to the behavioral change model \cite{li2011understanding}, early work of self-tracking tended to assume that reflection would simply occur by making the data available \cite{baumer2014reviewing}. Although some studies revealed challenges with data engagement, the challenges were mainly perceived as barriers to adoption or behavioral changes \cite{rapp2016personal, lazar2015we} and not themselves centrally focused for systematic examination. Other works focused on novel approaches to data presentation to address the challenges \cite{li2009beyond, mamykina2006investigating, frost2013supporting}, they tend to approach the challenges as a cognitive or analytic issues. As such, we still lack a clear understanding of how data engagement really occur in practice. More recent work has started to draw on reflection theories to better inform design \cite{slovak2017reflective, baumer2015reflective, ploderer2014social}, which renders better conceptual and systematic understandings of reflection, and new insights for design, e.g. supporting group discussions for inquiry \cite{baumer2015reflective}. However, these theories are being developed not for self-tracking technologies per se. Without technologies in the loop, they miss key issues central to  self-tracking technology's data engagement. 

 With the development of self-tracking technologies, and with increasingly more automatically-collected health data made easily available in our lives, it does not simply help to reduce the labor needed for tracking \cite{choe2014understanding}, but rather more fundamentally,  to change the very mode of data engagement in practice. What we highlight through the study is that the meaning of self-tracking is not simply a matter of having the data, nor analyzing the data, but a matter of situated practices of data engagement. As shown in the study, far from being simply an interaction between users and their data, data engagement is embedded in a web of an individual's intention for tracking, domain knowledge, technical properties, other users, and learning resources. As Kuutti put it: ``Practices are wholes, whose existence is dependent on the temporal interconnection of all these elements, and cannot be reduced to, or explained by, any one single element'' \cite{kuutti2014turn}. We argue that to understand the data engagement issues of self-tracking technologies, we should approach them as part of the whole tracking practice, a notion that assembles all these elements into a holistic unit, and does not reduce them to a mere cognitive analysis.

\section{Acknowledgments}
We would like to thank our participants for sharing their experiences. This work is supported by the National Key Research and Development Plan of China award number(s): 2016YFB1001200, the National Natural Science Foundation of China (NSFC) award number(s): 61672167, 61932007.

%%
%% The next two lines define the bibliography style to be used, and
%% the bibliography file.
\bibliographystyle{ACM-Reference-Format}
\bibliography{stress_tracking}

\end{document}